
%
%
%
\def\@{{\char'100}}

\long\def\abstract#1{\bigskip{\advance\leftskip by 2true cm
\advance\rightskip by 2true cm\eightpoint\centerline{\bf
Abstract}\everymath{\scriptstyle}\vskip10pt\vbox{#1}}\bigskip}
\long\def\resume#1{{\advance\leftskip by 2true cm
\advance\rightskip by 2true cm\eightpoint\centerline{\bf
R\'esum\'e}\everymath{\scriptstyle}\vskip10pt \vbox{#1}}}

\def\references{\bigbreak\centerline{\sc
References}\bigskip\nobreak\bgroup \baselineskip 13pt
\def\ref##1&{\leavevmode\hangindent16pt
\hbox to 14pt{\hss${}{##1.}$\ }\ignorespaces}
\parindent=0pt
\everypar={\ref}\par}
\def\endreferences{\egroup}
\long\def\authoraddr#1{\medskip{\baselineskip9pt\let\\=\cr
\halign{\line{\hfil{\Addressfont##}\hfil}\crcr#1\crcr}}}
\def\Subtitle#1{\medbreak\noindent{\Subtitlefont#1.} }
%
%
\newif\ifrunningheads
\runningheadstrue
\immediate\write16{- Page headers}
\headline={\ifrunningheads\ifnum\pageno=1\hfil\else\ifodd\pageno\rightheadline
\else\leftheadline\fi\fi\else\hfil\fi}
\def\rightheadline{\sc\hfil\RightHeadText\hfil}
\def\leftheadline{\sc\hfil\LeftHeadText\hfil}

\hyphenation{Harnad Neumann}
%
%
\immediate\write16{- Fonts "Small Caps" and "EulerFraktur"}
%
%
%

\let\sc=\tensmc
%
%
\font\teneuf=eufm10  \font\seveneuf=eufm7 \font\fiveeuf=eufm5
\newfam\euffam \def\gr{\fam\euffam\teneuf}
\textfont\euffam=\teneuf \scriptfont\euffam=\seveneuf
\scriptscriptfont\euffam=\fiveeuf
%

\def \wt {\widetilde}
\def \wh {\widehat}

\def \a {\alpha}
\def \b {\beta}
\def \d {\delta}
\def \e {\epsilon}

\def \l {\lambda}

\def \s {\sigma}

\def \ss {\subset}

\def \di {\partial}

\def\nchi{\hbox{\raise 2.5pt\hbox{$\chi$}}}
%
%

\def\nchi{\hbox{\raise 2.5pt\hbox{$\chi$}}}
%
%

%
%

%
%
\def\authorfont{\sc}
\font\eightrm=cmr8
\font\eightbf=cmbx8
\font\eightit=cmti8
\font\eightsl=cmsl8

\def\eightpoint{\let\rm=\eightrm \let\bf=\eightbf \let\it=\eightit
\let\sl=\eightsl \baselineskip = 9.5pt minus .75pt  \rm}

\font\titlefont=cmbx10 scaled\magstep2
\font\sectionfont=cmbx10
\font\Subtitlefont=cmbxsl10
\font\Addressfont=cmsl8
%
%
\def\Satz#1:#2\par{\smallbreak\noindent{\sc #1:\ }
{\sl #2}\par\smallbreak}
%
%
\immediate\write16{- Section headings}
\newcount\secount
\secount=0
\newcount\eqcount
\outer\def\section#1.#2\par{\global\eqcount=0\bigbreak
\ifcat#10
 \secount=#1\noindent{\sectionfont#1. #2}
\else
 \advance\secount by 1\noindent{\sectionfont\number\secount. #2}
\fi\par\nobreak\medskip}
%
%
\immediate\write16{- Automatic numbering}
\catcode`\@=11
\def\adv@nce{\global\advance\eqcount by 1}
\def\unadv@nce{\global\advance\eqcount by -1}
\def\nextnumber{\adv@nce}
%
%
\newif\iflines
\newif\ifm@resection
\def\onesec{\m@resectionfalse}
\def\moresec{\m@resectiontrue}
\moresec
\def\eq{\global\linesfalse\eq@}
\def\eqn{\global\linestrue&\eq@}
\def\nosubind@x{\global\subind@xfalse}
\def\newsubind@x{\ifsubind@x\unadv@nce\else\global\subind@xtrue\fi}
\newif\ifsubind@x
\def\eq@#1.#2.{\adv@nce
 \if\relax#2\relax
  \edef\loc@lnumber{\ifm@resection\number\secount.\fi
  \number\eqcount}
  \nosubind@x
 \else
  \newsubind@x
  \edef\loc@lnumber{\ifm@resection\number\secount.\fi
  \number\eqcount#2}
 \fi
 \if\relax#1\relax
 \else
  \expandafter\xdef\csname #1@\endcsname{{\rm(\loc@lnumber)}}
  \expandafter
  \gdef\csname #1\endcsname##1{\csname #1@\endcsname
  \ifcat##1a\relax\space
  \else
   \ifcat\noexpand##1\noexpand\relax\space
   \else
    \ifx##1$\space
    \else
     \if##1(\space
     \fi
    \fi
   \fi
  \fi##1}\relax
 \fi
 \eq@@{\loc@lnumber}}
\def\eq@@#1{\iflines \else \eqno\fi{\rm(#1)}}
\def\m@th{\mathsurround=0pt}
%
%
\def\display#1{\null\,\vcenter{\openup1\jot
\m@th
\ialign{\strut\hfil$\displaystyle{##}$\hfil\crcr#1\crcr}}
\,}
\newif\ifdt@p
\def\@lign{\tabskip=0pt\everycr={}}
\def\displ@y{\global\dt@ptrue \openup1 \jot \m@th
 \everycr{\noalign{\ifdt@p \global\dt@pfalse
  \vskip-\lineskiplimit \vskip\normallineskiplimit
  \else \penalty\interdisplaylinepenalty \fi}}}
%
%
\def\displayno#1{\displ@y \tabskip=\centering
 \halign to\displaywidth{\hfil$
\@lign\displaystyle{##}$\hfil\tabskip=\centering&
\hfil{$\@lign##$}\tabskip=0pt\crcr#1\crcr}}
%
%
\def\cite#1{{[#1]}}
\catcode`\@=\active
%
%
\magnification=\magstep1
\hsize= 6.75 true in
\vsize= 8.75 true in
%
%
\def\RightHeadText{Hyperspherical Harmonics and Bethe Ansatz}
\def\LeftHeadText{J. Harnad and P. Winternitz}
%
%
\rightline{CRM-2174 (1994) \break}
\bigskip
\centerline{\titlefont Hyperspherical Harmonics, Separation of Variables}
\centerline{\titlefont and the Bethe Ansatz}
\bigskip
\centerline{\authorfont J.~Harnad}
\authoraddr
{Department of Mathematics and Statistics, Concordia University\\
7141 Sherbrooke W., Montr\'eal, Qu\'ebec, Canada H4B 1R6, {\rm \eightpoint
and}\\
Centre de recherches math\'ematiques, Universit\'e de Montr\'eal\\
C.~P.~6128, succ. centre ville, Montr\'eal, Qu\'ebec, Canada H3C 3J7\\
{\rm \eightpoint e-mail}: harnad\@alcor.concordia.ca \ {\rm \eightpoint or}\
harnad\@mathcn.umontreal.ca}
\bigskip
\centerline{\authorfont  P.~Winternitz}
\authoraddr
{Centre de recherches math\'ematiques, Universit\'e de Montr\'eal\\
C.~P.~6128 succ. centre ville, Montr\'eal, Qu\'ebec, Canada H3C 3J7\\
{\rm \eightpoint e-mail}: wintern\@ere.umontreal.ca}
\bigskip
\abstract{The relation between solutions to Helmholtz's equation on the sphere
$S^{n-1}$  and the $[{\gr sl}(2)]^n$ Gaudin spin chain is clarified. The joint
eigenfuctions of the Laplacian and a complete set of commuting second order
operators suggested by the $R$--matrix approach to integrable systems, based on
the
loop algebra $\wt{sl}(2)_R$, are found in terms of homogeneous polynomials in
the
ambient space. The relation of this method of determining a basis of harmonic
functions on $S^{n-1}$ to the Bethe ansatz approach to integrable systems is
explained.} \bigskip \baselineskip 14 pt
\noindent{\sectionfont Introduction}\medskip\nobreak
   In some recent papers on separation of variables in the Hamilton-Jacobi and
Schr\"odinger equations \cite{1--3}, it has been asserted that free
motion, both classical and quantum, on the sphere $S^{n-1}$ and on the negative
constant curvature hyperboloid $H^{n-1}$ is equivalent to the $[{\gr sl}(2)]^n$
Gaudin spin chain \cite{4,5}. There is, in fact, a close relationship
between these systems, which may best be seen through the method of moment map
embeddings in loop algebras \cite{6--9}, but they actually differ
significantly at the quantum level. The purpose of this work is, first of all,
to
clarify what this relationship is and, secondly, to show how an approach that
is
related to the functional Bethe ansatz \cite{10} for spin chains leads very
simply to a basis of harmonic functions on $S^{n-1}$  generalizing that
provided,
for the case $n=3$, by the Lam\'e polynomials \cite{11,12}.

  Briefly, the difference between the Laplacian on $S^{n-1}$ and the
$[{\gr sl}(2)]^n $ spin chain system is that, while the former does belong to a
commuting family which formally is of the same type as the Gaudin systems, the
permissible values of the ${\gr sl}(2)$ Casimir invariants are different for
the
two cases. Hence, whereas the classical phase spaces may be simply related, the
quantum Hilbert space, joint eigenstates and spectrum are quite different.
Moreover, the mapping which determines this correspondence is not $1$--$1$, but
rather involves a quotient by the group ${\bf Z}_2^n$ of reflections in the
coordinate planes. As a consequence, the Laplacian on $S^{n-1}$, together with
the
associated commuting family of second order operators provided by the loop
algebra
framework, is diagonalized not on a unique highest weight module of $[{\gr
sl}(2)]^n$, as in the Gaudin systems, but rather on the sum of $2^n$ invariant
subspaces, each characterized by distinct ${\bf Z}_2^n$ transformation
properties
and containing its own highest weight vector. Essentially, the $[{\gr
sl}(2)]^n$
spin chains are defined on tensor products of irreducible discrete series
representations of ${\gr sl}(2)$, while  the Laplacian on $S^{n-1}$ involves
products of the direct sum of the two lowest metaplectic representations. Thus
there are, roughly speaking, $2^n$ times as many harmonic functions on
$S^{n-1}$ as
joint eigenstates in an irreducible Gaudin system.

   In the following, these joint eigenfunctions will be constructed through
a separation of variables technique originally developed in the loop algebra
approach to classical integrable systems \cite{13,14}. In the quantum
setting this leads directly to the functional Bethe ansatz eigenstates. The
classical case is already amply treated in \cite{7--9,13,14}, so we proceed
directly to the quantum case here. The canonical quantization approach leading
to
the construction used below is detailed in \cite{15}. Similar formulations may
be
found in \cite{1,2,10,16}. \bigskip
\section 1. Quantum Moment Map Construction and Commuting Invariants
\medskip
\Subtitle {1a. $\wt{sl}(2)_R$ Loop Algebra Representation and Commuting
Invariants}

    Let $(x_1, \dots x_n)$ be the Cartesian coordinates in ${\bf R}^n$ and
$(\mu_1, \dots \mu_n)$ a set of $n$ real numbers. Define the operators
$$
\eqalignno{
e_i&:= {1\over 2}\left({\di^2\over \di x_i^2}-
{\mu_i^2\over x_i^2}\right)\eqn eii.a. \cr
f_i&:= {1\over2}x_i^2   \eqn fii.b.  \cr
h_i&:={1\over 2}\left(x_i{\di \over \di x_i} +{1\over 2}\right),
\qquad i=1, \dots n, \eqn hii.c.}
$$
which satisfy the commutation relations
$$
[h_i, \ f_i]=f_i, \quad [h_i, \ e_i] = - e_i, \quad [e_i,\ f_i] = 2h_i,
\eq sl2i..
$$
and hence determine $n$ representations of ${\gr sl}(2)$.
The constants $\{\mu_1, \dots \mu_n\}$ are related to the values of the
${\gr sl}(2)$ Casimir invariants by
$$
h_i^2 -{1\over 2}(e_i f_i+f_i e_i)= {1\over 4}\left(\mu_i^2-{3\over 4}\right).
\eq sl2Casimir..
$$

   We shall only be concerned here with quadratic combinations of these
operators
that may consistently be restricted to the  unit sphere $S^{n-1}\ss {\bf R}^n$.
Pick a further set $\{\a_1, \dots \a_n\}$ of real constants, all distinct,
and ordered with increasing values. Define now the following
operator--valued rational functions of the ``loop'' parameter $\l$
$$
\eqalignno{
e(\l)&:= \sum_{i=1}^n {e_i\over \l-\a_i}   \eqn elam.a. \cr
f(\l)&:= \sum_{i=1}^n {f_i\over \l-\a_i}   \eqn flam.b. \cr
h(\l)&:= a + \sum_{i=1}^n {h_i\over \l-\a_i},  \eqn hlam.c.}
$$
\nextnumber
where $a$ is a real constant. These satisfy the well-known commutation
relations
for the loop algebra $\wt{sl}(2)_R$  defined with respect to a rational
classical
$R$--matrix structure \cite{6,17}:
 $$
\eqalignno{
[h(\l), \ e(\mu)] &={e(\l)-e(\mu) \over \l - \mu}  \eqn hecom.a.\cr
[h(\l), \ f(\mu)] &=-{f(\l)-f(\mu) \over \l - \mu}  \eqn hfcom.b. \cr
[e(\l), \ f(\mu)] &=-2{h(\l)-h(\mu) \over \l - \mu},   \eqn efcom.c.}
$$

  It follows that the  coefficients of the following operator--valued rational
function
$$
\Delta(\l):=h^2(\l) -{1\over 2} \left(e(\l) f(\l) +f(\l) e(\l)\right)
\eq Deltalam..
$$
commute amongst themselves:
$$
[\Delta(\l),\ \Delta(\mu)]=0,  \qquad \forall \ \l, \mu.  \eq Deltacomm..
$$
Expanding $\Delta(\l)$ in partial fractions gives
$$
\Delta(\l) = a^2 + \sum_{i=1}^n{H_i\over \l -\a_i}
+K(\l) + L^2(\l)+ L'(\l),
\eq Deltapfrac..
$$
where
$$
\eqalignno{
K(\l)&:={1\over 4}\sum_{i=1}^n{\mu_i^2\over (\l-\a_i)^2}  \eqn K_def.a.\cr
L(\l)&:={1\over 4}\sum_{i=1}^n{1\over \l-\a_i}     \eqn Ldef.b.}
$$
and $\{H_i\}_{i=1,\dots n}$ are a set of commuting second order
differential operators defined by
$$
H_i:={1\over 4}\sum_{{j=1\atop j\neq i}}^n{-L_{ij}^2
+\mu_i^2{x_j^2\over x_i^2} +\mu_j^2{x_i^2\over x_j^2}\over \a_i -\a_j}
+a\left(x_i {\di\over \di x_i} +{1\over 2}\right),  \eq Hi..
$$
where
$$
L_{ij}:=x_i{\di\over \di x_j} -x_j{\di\over \di x_i}  \eq Lij..
$$
is the generator of rotations in the $(ij)$ plane. For $a\neq 0$, the $H_i$'s
are linearly independent and their sum is
$$
\sum_{i=1}^n H_i = a (D + {n\over 2}),   \eq sumHi..
$$
where
$$
D:= \sum_{i=1}^n x_i{\di\over \di x_i}  \eq Dop..
$$
is the Euler homogeneity operator. For $a=0$, this sum vanishes, so only $n-1$
of
the operators are independent. But since they still all commute with the Euler
operator $D$, this can be adjoined as the $n$th independent commuting invariant
in this case.

  Contained within this framework, for particular values of the constants
$\{a, \mu_1, \dots \mu_n\}$, are both the Laplacian on the sphere $S^{n-1}$ and
the Gaudin $[{\gr sl}(2)]^n$ spin chains. It is important to note, however,
that
they correspond to {\it different} values of the Casimir invariants for the
various
${\gr sl}(2)$ representations. To obtain the Laplacian on $S^{n-1}$, we set $a$
and all the $\mu_i$'s equal to zero. Since the $H_i$'s are defined in terms
of the rotation generators $L_{ij}$, they may be restricted consistently to
$S^{n-1}$. Forming the sum
$$
H_0:= 4\sum_{i=1}^n\a_iH_i =-{1\over 2}\sum_{i,j=1}^n L_{ij}^2  \eq
H0alphasum..
$$
and restricting to the unit sphere $S^{n-1}$ gives the Laplacian
$$
H_0|_{S^{n-1}}=-\Delta_{S^{n-1}}.  \eq HLaplac..
$$
More generally, if the $\mu_i$'s are nonvanishing, we have
$$
H:= 4\sum_{i=1}^n \a_i H_i =-{1\over 2}\sum_{i,j=1}^n L_{ij}^2 +
 \sum_{i=1}^n x_i^2 \sum_{j=1}^n{\mu_j^2\over x_j^2} -
 \sum_{i=1}^n\mu_i^2,
 \eq Halphasum..
$$
and restricting to $S^{n-1}$ gives
$$
H|_{S^{n-1}}= - \Delta_{S^{n-1}} +\sum_{i=1}^n {\mu_i^2\over x_i^2}
- \sum_{i=1}^n\mu_i^2,.
\eq HS..
$$
This corresponds to a degenerate case of the quantum Rosochatius system
\cite{15,18}, in which the harmonic oscillator potential is absent. (It may
also
be obtained as a reduction of the free system on a sphere of dimension $2n-1$
under the action of the maximal torus in the isometry group $SO(2n)$
(cf. \cite{19}). The other operators $H_i$ in the commuting family may
similarly be restricted to define commuting operators on $S^{n-1}$.
\medskip
\Subtitle {1b. Relation to the $[{\gr sl}(2)]^n$ Gaudin Chain}

  To obtain the $[{\gr sl}(2)]^n$ Gaudin spin chain, we must make a change of
representation. Instead of considering square--integrable functions on
${\bf R}^n$ or $S^{n-1}$, we must reinterpret the operators $\{e_i, f_i, h_i\}$
entering in the definitions \elam--\hlam, \Deltalam as acting
upon a highest weight module of $[{\gr sl}(2)]^n$ of the type
$$
{\cal H}=\otimes_{i=1}^n{\cal H}_{l_i},  \eq Hilbert..
$$
where the individual factors ${\cal  H}_{l_i}$ in the tensor product are
generated by application of the ladder operator $f_i$ to a unique highest
weight
vector $|0\rangle_{l_i}$, the kernel of the operator $e_i$. (We make no
notational
distinction between an operator ${\cal O}$ acting on ${\cal H}_{l_i}$ and its
extension $I\otimes I\otimes \cdots \otimes{\cal O}\otimes \cdots \otimes I$
acting on ${\cal H}$.)

  Note that the space of square-integrable smooth functions on ${\bf R}^n$,
with the representation \eii--\hii, will not do for this purpose, since the
kernel of the operator \eii is two dimensional. However, ignoring for the
present the question of normalizability, we may still consider the space
obtained by application of polynomials in the operators $f_i$ to any
specifically
chosen vector $|0\rangle_{l_i}$ within the kernel of each $e_i$ . This kernel
is
of the form
$$
{\cal K}_i=\{a_ix_i^{\b_i} + b_i x_i^{-(\b_i-1)}\},  \eq kerneli..
$$
where
$$
\b_i(\b_i-1)=\mu_i^2.  \eq betai..
$$
For $\mu_i\neq 0$, we may choose the unique positive root of this quadratic
equation and identify, up to normalization
$$
|0\rangle_{l_i} \sim x^{\b_i}.  \eq 0li..
$$
If the domain of definition is ${\bf R}^n$, to obtain single--valued functions,
the $\b_i$'s must be chosen to have integer values.
Then ${\cal H}_{l_i}$ consists of functions of the form
$ p (x_i^2)x_i^{\b_i}$, where $p$ is a polynomial in $x_i^2$, and
${\cal H}$ consists of functions of the form ${\cal P}(x_1^2, \dots
x_n^2)\prod_{i=1}^n x_i^{\b_i}$, where ${\cal P}$ is a polynomial in its $n$
arguments $\{x_1^2, \dots x_n^2\}$.
In the case where the $\mu_i$'s vanish, however, which is the one relevant to
the
Laplacian on $S^{n-1}$, each of the $\b_i$'s may have the  {\it two} possible
values
$$
\b_i=0  \quad {\rm or} \quad \b_i =1,  \eq beta01..
$$
and we obtain distinct highest weight representations for each choice of the
$\b_i$'s.

 To obtain the ${\gr sl}(2)$ analogue of the highest weight representations
used in
\cite{10}, we must conjugate the operators $\{e_i, f_i, h_i\}$ by the factor
$x_i^{\b_i}$ and re-express the result in terms of the new variable
$$
z_i:={1\over 2}x_i^2.  \eq zi..
$$
Thus, we define
$$
\eqalignno{
\wh{e}_i&:= x_i^{-\b_i} e_i x_i^{\b_i} = z_i{\di^2\over \di z_i^2}
+2l_i {\di \over \di z_i}  \eqn ehati.a. \cr
 \wh{f}_i&:= x_i^{-\b_i} f_i x_i^{\b_i} = z_i \eqn fhati.b. \cr
\wh{h}_i&:= x_i^{-\b_i} h_i x_i^{\b_i} = z_i{\di\over \di z_i} +l_i,
\eqn hhati.c. }
$$
where
$$
l_i:={\b_i\over 2}+{1\over 4}.  \eq lsubi..
$$
This is the ${\gr sl}(2)$ analogue of the representation used in \cite{10} to
describe the ${\gr su}(2)$ Gaudin spin chain in functional terms
(cf. \cite{20,21}). The highest weight state $|0\rangle$ in this representation
is
just a constant, and  a basis for the corresponding space $\wh{\cal H}$
generated by
application of the $\wh{f}_i$'s is given by the homogeneous polynomials in the
variables $\{z_1, \dots z_n\}$.

  Within such representations, the $\b_i$'s need not have integer values. For
any
set of positive values for the $l_i$'s, a scalar product may be defined by:
$$
\langle z_1^{j_1}\cdots z_n^{j_n},\ z_1^{k_1}\cdots z_n^{k_n}\rangle :=
\prod_{i=1}^n\d_{j_i k_i}C_{l_i,j_i}, \quad j_i, k_i \in {\bf N},
\eq scalarprod..
$$
where
$$
C_{l_i, j_i}:=(j_i)!\prod_{j=1}^{j_i}(2l_i +j-1) \quad {\rm for}\quad j_i\ge1,
\quad \  C_{l_i,0}:=1.  \eq clj..
$$
With respect to this scalar product, the operators $\wh{h}_i$ are hermitian,
while
$\wh{e}_i$ and $\wh{f}_i$ are mutually hermitian conjugate, so we are really
dealing with unitary representations of the real form ${\gr su}(1,1)$. It
follows
that, for the case of integer $\b_i$'s,  the operators $\wh{H}_i$ obtained by
conjugating the $H_i$'s by the factor $\prod_{i=1}^n x_i^{\b_i}$ are all
hermitian.
Within such representations, these may be viewed as the Hamiltonians of the
$[{\gr sl}(2)]^n$ Gaudin spin chain \cite{4,10}. However, it is only  integer
or
half--integer $l_i$'s that define discrete series representations; integer
$\b_i$'s
give quarter--integer $l_i$'s, corresponding instead metaplectic
representations
\cite{22}.

  Moreover, for the case when all the $\mu_i$'s vanish, we must replace the
factors  ${\cal H}_{l_i}$ in \Hilbert, by the direct sum  ${\cal H}_{1 \over
4}\oplus{\cal H}_{3 \over 4}$ of the two lowest metaplectic representations.
Therefore, we do not have a unique highest weight representation. Expressed in
terms of the original Cartesian coordinates $\{x_1, \dots x_n\}$, to each
invariant
subspace in the $z_i$ representation consisting of  homogeneous polynomials in
$\{z_1, \dots z_n\}$ of degree $p$, there are associated $2^n$ different
possible
invariant subspaces, consisting of polynomials in  $\{x_1, \dots x_n\}$ of
degree
\hbox{$2p+\sum_{i=1}^n\b_i$}, one for each choice of the binary sequence
$\{\b_1,
\dots \b_n\}$. These are not only invariant under the representation of $[{\gr
sl}(2)]^n$ defined by  \eii--\hii, but also under the group ${\bf Z_2}^n$ of
reflections in the coordinate planes, and each has its own highest weight
vector.
\bigskip

\section 2. Determination of Joint Eigenfunctions; Separation of Variables
\medskip
\Subtitle {2a. Separating Coordinates and Irreducible Subspaces}

  We now introduce a coordinate system that is specially adapted to the
simultaneous diagonalization of the commuting operators $H_i$ introduced
above (cf. \cite{13,14}), the sphero-conical coordinates $\{r, \l_1,\dots
\l_{n-1}\}$, defined by
$$
\sum_{i=1}^{n}{x_i^2\over \l-\a_i}={r^2 Q(\l)\over a(\l)},  \eq..
$$
where
$$
\eqalignno{
Q(\l)&:=\prod_{\mu=1}^{n-1}(\l-\l_\mu)   \eqn..  \cr
a(\l)&:=\prod_{i=1}^n(\l-\a_i), \eqn..}
$$
with the $\l_\mu$'s chosen in the range:
$$
\a_1<\l_1<\a_2< \cdots <\l_{n-1}<\a_n.   \eq..
$$
The functions $x_i^2$ are given in terms of the coordinates
$\{r, \l_1, \dots \l_{n-1}\}$ by
$$
x_i^2 =  {2 r^2\prod_{\mu=1}^{n-1}(\a_i -\l_\mu) \over a'(\a_i) }, \eq
xisquared..
$$
and hence are linear combinations of the elementary symmetric invariants
$$
\s_k := \sum_{i_1<i_2< \cdots <i_k}\l_{i_1} \cdots \l_{i_k},
\qquad k=0, \dots  n-1.  \eq symminvars..
$$

   Define the operator--valued polynomial $\wh{\cal P}(\l)$ by
$$
{\wh{\cal P}(\l)\over a(\l)} := \sum_{i=1}^n {H_i\over \l-\a_i}. \eq..
$$
Since the coefficients of $\wh{\cal P}(\l)$ are just linear combinations of the
$H_i$'s, simultaneous diagonalization of the latter is equivalent to
diagonalization of $\wh{\cal P}(\l)$  on functions that do not depend on the
parameter $\l$. Noting that the operator $h(\l)$, evaluated at $\l=\l_\mu$ is
just
$$
h(\l)|_{\l=\l_\mu} = {\di \over \di \l_\mu} + L(\l_\mu) + a,  \eq..
$$
and that $f(\l_\mu)$ vanishes at $\l=\l_\mu$, it follows from \Deltalam,
\Deltapfrac that
$$
\wh{\cal P}(\l)|_{\l=\l_\mu}= a(\l_\mu)\left( {\di^2\over \di \l_\mu^2}
   +2(a + L(\l_\mu)){\di \over \di \l_\mu}+2aL(\l_\mu) - K(\l_\mu)\right).
\eq Phat..
$$
Using also the fact that the $\l^{n-1}$ coefficient of $\wh{\cal P}(\l)$
is given by \sumHi,  Lagrange interpolation shows that $\wh{\cal P}(\l)$ may be
expressed in terms of the $\{r, \l_1, \dots \l_{n-1}\}$ coordinates as
$$
\eqalignno{
\wh{\cal P}(\l) = &\sum_{\mu=1}^{n-1} {Q(\l) a(\l_\mu)\over
Q'(\l_\mu)(\l-\l_\mu)}
\left[{\di^2\over \di \l_\mu^2}
   +2(a + L(\l_\mu)){\di \over \di \l_\mu} +2aL(\l_\mu) - K(\l_\mu)\right]
 \cr
 & +aQ(\l)\left(r{\di\over \di r }+ {n\over 2}\right).  \eqn Phatlam..}
$$

   Now assume that the $\mu_i$'s are given by \betai, with the $\b_i$'s all
nonnegative integers. Let ${\cal H}^p(\b_1, \dots ,\b_n)$ denote the space of
homogeneous polynomials of the form
$$
{\cal P}(x_1^2, \dots, x_n^2) \prod_{i=1}^n x_i^{\b_i},  \eq..
$$
where ${\cal P}(x_1^2, \dots, x_n^2)$ is a homogeneous polynomial of degree $p$
in its arguments. In the following, we must distinguish between the case when
all
the $\mu_i$'s vanish, for which there are $2^n$ different such spaces for each
$p$, and the case when they do not. For the former, the operators  $H_i$ are
well-defined for all values of the $x_i$'s. Restricting to the  unit sphere
$S^{n-1}$, the resulting operators are still well defined, because the Euler
operator
$$
 D=r{\di\over \di r }  \eq..
$$
entering in \Phatlam takes a fixed value on each ${\cal H}^p(\b_1, \dots,
\b_n)$.
For the case of nonvanishing $\mu_i$'s, the operators $H_i$ are singular on
the hyperplanes $x_i=0$ whenever $\mu_i\neq 0$. However, restricting to the
space ${\cal H}^p(\b_1, \dots, \b_n)$, the $H_i$'s become regularized through
the
conjugation by $x_i^{\b_i}$ that was used in defining the operators
$\{\wh{e}_i,
\wh{f}_i, \wh{h}_i\}$ in \ehati--\hhati. We see that the Hamiltonian operators
\Halphasum and \HS  contain potentials that give rise to an infinite repulsion
away from these coordinate hyperplanes, and hence the eigenfunctions must
vanish to
appropriate order at these planes.
\medskip
\Subtitle {2b. Determination of Eigenfunctions and the Functional Bethe Ansatz}

  We seek joint eigenfunctions $\Psi$, satisfying
$$
H_i \Psi= E_i \Psi \eq..
$$
or, equivalently
$$
\wh{\cal P}(\l)\Psi = E(\l)\psi,  \eq Schrodeq..
$$
where the polynomial $E(\l)$ is defined by
$$
{E(\l)\over a(\l)} = \sum_{i=1}^n {E_i\over \l-\a_i}.  \eq..
$$
Since the Euler operator $D$ and the generators of the group ${\bf Z}_2^n$
of reflections in the coordinate planes all commute with the $H_i$'s, these may
be simultaneously diagonalized. This, together with the regularization
requirement
in the case when some $\mu_i$'s are nonzero, implies that each joint
eigenfunction $\Psi$ belongs to one of the spaces ${\cal H}^p(\b_1, \dots,
\b_n)$, and the Euler operator in \Phatlam may be replaced by its eigenvalue
$2p+\sum_{i=1}^n \b_i$. The resulting operators may consistently be restricted
to
the unit sphere $S^{n-1}$. Expressed in terms of the $\{r, \l_1, \dots,
\l_\mu\}$
coordinates, $\Psi$ is therefore of the form
$$
\Psi(r, \l_1, \dots, \l_{n-1})=
r^{(2p+{\scriptscriptstyle\sum_{i=1}^n}\b_i)}\psi(\l_1, \dots ,\l_{n-1})
\prod_{i=1}^n \prod_{\mu=1}^{n-1} (\l_\mu -\a_i)^{\b_i\over 2},  \eq Psidef..
$$
where $\psi(\l_1, \dots ,\l_{n-1})$ is a  homogeneous polynomial of degree $p$
in the symmetric invariants $\{\s_0, \dots \s_{n-1}\}$, and hence, of degree
$\le p(n-1)$ in the $\l_\mu$'s. (For odd values of $\b_i$, the apparent sign
ambiguity in the factors $(\l_\mu -\a_i)^{\b_i\over 2}$ is resolved by
identifying
the sign of the product over $\mu$ with that of $x_i^{\b_i}$.) The eigenvalue
equation \Schrodeq then reduces to
$$
\eqalignno{
\sum_{\mu=1}^{n-1}{a(\l_\mu)\over Q'(\l_\mu)(\l-\l_\mu)}
&\left[{\di^2 \psi\over \di \l_\mu^2}
+2(a +\Lambda(\l_\mu)){\di \psi \over \di\l_\mu} +
{1\over 4}\sum_{j, i=1 \atop j\neq i}^n {2\b_i\b_j +\b_i +\b_j\over
(\a_i-\a_j)(\l_\mu -\a_i)}\psi
+2a \Lambda(\l_\mu)\psi\right] \cr
&+a\left(2p + \sum_{i=1}^n\b_i + {n\over 2}\right)\psi
={E(\l)\over Q(\l)}\psi,  \eqn Schrodredeq..}
$$
where
$$
\Lambda(\l) := {1\over 2}\sum_{i=1}^n{\b_i +{1\over 2}\over \l -\a_i}.
\eq Lambdadef..
$$
Equating residues at each $\l=\l_\mu$, this is equivalent to
$$
{\di^2 \psi\over \di \l_\mu^2}
+2(a +\Lambda(\l_\mu)){\di \psi \over \di\l_\mu}+2a \Lambda(\l_\mu)\psi +
{1\over 4}\sum_{j, i=1 \atop j\neq i}^n {2\b_i\b_j +\b_i +\b_j\over
(\a_i-\a_j)(\l_\mu -\a_i)}\psi
={E(\l_\mu)\over a(\l_\mu)}\psi.  \eq Schrodsep..
$$
By uniqueness (up to normalization) of polynomial solutions of equations of
the type  \Schrodsep, $\psi$ must have the factorized form
$$
\psi(\l_1, \dots , \l_{n-1}) =\prod_{\mu=1}^{n-1} q(\l_\mu), \eq..
$$
where $q(\l)$ is a polynomial of degree $\le p$ satisfying the equation
$$
a(\l)q''(\l) +2 b(\l)q'(\l) + c(\l) q(\l)=0,  \eq HeineStieltjes..
$$
with polynomial coefficients $b(\l)$, $c(\l)$ defined by
$$
\eqalignno{
b(\l) &:= a\left(\l)(a + \Lambda(\l)\right)   \eqn blambda.a. \cr
c(\l) &:= a(\l)\left(2a \Lambda(\l) +
{1\over 4}\sum_{j, i=1 \atop j\neq i}^n {2\b_i\b_j +\b_i +\b_j\over
(\a_i-\a_j)(\l -\a_i)}\right) -E(\l).    \eqn clambda.b. }
$$

  For the case $a=0$, the number of distinct polynomials of degree $p$
satisfying
an equation of the type \HeineStieltjes is known, by the Heine-Stieltjes
theorem
\cite{23}, to be $\pmatrix{n+p-2 \cr p}$. This may also be seen as follows.
Note
that the operators $H_i$ are all hermitian with respect to the scalar product
on
${\cal H}^p(\b_1, \dots, \b_n)$ determined from \scalarprod through the
isomorphism between the space of homogeneous polynomials ${\cal P}(z_1, \dots,
z_n)$ of degree $p$ and ${\cal H}^p(\b_1, \dots, \b_n)$ given by multiplication
by the factor $\prod_{i=1}^n x_i^{\b_i}$. Their joint eigenfunctions therefore
provide a basis for this space, which is of dimension  $\pmatrix{n+p-1 \cr p}$.
Each such joint eigenfunction corresponds to a unique polynomial solution of
the
equation \HeineStieltjes for some set of eigenvalues $\{E_1, \dots, E_n\}$.
These
include not only polynomials of degree $p$, but also those of all lower degree.
The corresponding eigenfunctions are obtained from an eigenfunction in some
lower
degree space  ${\cal H}^{p'}(\b_1, \dots, \b_n)$,\quad $p'<p$ by multiplication
by
$r^{2(p-p')}$. On the sphere, therefore, the number of eigenfunctions in
${\cal H}^{p}(\b_1, \dots, \b_n)|_{S^{n-1}}$ which do not coincide with one in
some lower degree space ${\cal H}^{p'}(\b_1, \dots, \b_n)|_{S^{n-1}}$ is
$$
\pmatrix{n+p-1\cr p}-\pmatrix{n+p-2\cr p-1}=\pmatrix{n+p-2\cr p}.  \eq..
$$
For $a\neq0$, mutiplication of the eigenfunctions in
${\cal H}^{p'}(\b_1, \dots, \b_n)$ by $r^{2(p-p')}$ does not produce
eigenfunctions in ${\cal H}^{p}(\b_1, \dots, \b_n)|_{S^{n-1}}$, due to the last
term in \Phat. Therefore the eigenfunctions in
${\cal H}^{p}(\b_1, \dots, \b_n)|_{S^{n-1}}$ must include those of lower
degree,
giving a total of $\pmatrix{n+p-1 \cr p}$.

  To actually compute the eigenfunctions, we express $q(\l)$ in factorized form
$$
q(\l)=\prod_{b=1}^p(\l - v_b),  \eq..
$$
and substitute in \HeineStieltjes.  Dividing by $a(\l)$, the resulting rational
function vanishes if the singular parts at each of the poles
$\{\l=\a_i\}_{i=1, \dots n}$ and $\{\l=v_a\}_{a=1, \dots p}$ vanish. This gives
the
secular equations determining the nodes $\{v_a\}_{a=1, \dots p}$ as
$$
2\sum_{c=1\atop c\neq b}^p {1\over v_b-v_c}+\sum_{i=1}^n{\b_i+{1\over 2} \over
v_b-\a_i}  + 2a =0,  \quad b=1, \dots, p, \eq BGva..
$$
and the eigenvalues $\{E_i\}_{i=1,\dots n}$ as
$$
E_i = \left(\b_i +{1\over 2}\right)\sum_{b=1}^p{1\over \a_i - v_b} +
{1\over 4} \sum_{j=1\atop j\neq i}^n {2\b_i\b_j+\b_i+\b_j\over \a_i-\a_j}
+ a\left(\b_i+{1\over2}\right). \eq BGEi..
$$

  These are the same as the Bethe-Gaudin equations occurring in the  Bethe
ansatz
solutions to the ${\gr sl}(2)$ spin chain \cite{4,5,10}. However, for the case
of
the Laplacian $\Delta|_{s^{n-1}}$, they differ in interpretation, since the
quantities  $\{l_i={\b_i\over 2}+{1\over 4}\}_{i=1, \dots n}$ do not take
unique
integer values, but rather all possible  $2^n$ combinations of the values
${1\over
4}$ or ${3\over 4}$. The resulting eigenvectors, within normalization, are
precisely of the Bethe ansatz form:
$$
\Psi=\prod_{a=1}^p f(v_a) |0;\b_1, \dots,\b_n\rangle, \eq BetheState..
$$
where
$$
|0;\b_1, \dots,\b_n\rangle := \prod_{i=1}^n x_i^{\b_i}  \eq betavacuum..
$$
denotes the highest weight vector in the ${\bf Z}_2^n$--invariant
subspace ${\cal H}^p(\b_1, \dots, \b_n)$.

  Summing over the eigenvalues of the $H_i$'s to obtain that of $H$  in
\Halphasum gives
$$
4\sum_{i=1}^n\a_iE_i = (2p +\sum_{i=1}^n\b_i)(2p +\sum_{i=1}^n\b_i +n-2)
+\sum_{i=1}^n(\b_i-\b_i^2)
+4a\left(2\sum_{b=1}^p v_b +\sum_{i=1}^n\a_i\left(\b_i+{1\over2}\right)\right)
\eq.. $$
For the case when $a=0$ and each $\b_i$ is either $0$ or $1$, this
reduces to the usual formula \cite{24} for the eigenvalues $E_{n,l}$ of the
Laplacian on $S^{n-1}$
$$
E_{n,l}= -l(l+n-2),  \eq..
$$
expressed in terms of the total degree of homogeneity
$$
l:= 2p + \sum_{i=1}^n\b_i.   \eq..
$$

   To count the total number of eigenfunctions in this case with a given degree
of
homogeneity $l$,  let $\e=0$ or $1$, for $l$ even or odd, respectively, and
$$
m:={\rm min}\left(\left[{n-\e\over2}\right],\left[{l\over 2}\right]\right).
\eq..
$$
Summing over the number of eigenfunctions for each combination of values of $p$
and $\b_i$'s giving the same homogeneity $l$, the total number is
$$
a_{n,l} =\sum_{k=0}^m\pmatrix{n+\left[l\over 2\right]-2-k\cr
n-2}\pmatrix{n\cr2k+\e}
 ={(n+2l-2)(n+l-3)!\over l! (n-2)!},  \eq..
$$
which is the number of homogeneous polynomials of degree $l$ which may not be
obtained by multiplying those of degree $l-2$ by $r^2$. This agrees with
the usual count for the number of harmonic functions on $S^{n-1}$ obtained,
e.g.,
through the use of Gel'fand--Tseitlin bases \cite{24}.
\bigskip
\section 3.Conclusions and Discussion
\medskip

   We have seen that a basis of harmonic functions on $S^{n-1}$ may be obtained
in
terms of homogeneous polynomials that are simultaneous eigenfunctions of the
complete
set of commuting second order operators $\{H_1, \dots , H_n\}$ provided by the
$\wt{\gr sl}(2)_R$ loop algebra framework, as well as of the generators of the
group ${\bf Z}_2^n$ of reflections in the coordinate planes. This gives a
natural
generalization of the harmonic functions on $S^2$ provided by the Lam\'e
polynomials \cite{11,12}.  More generally, the same method determines the joint
eigenfunctions of the systems associated to the operator \Halphasum for
$\mu_i=\b_i(\b_i-1)$ with $\b_i$'s any nonnegative integers. These
eigenfunctions
provide bases for the spaces  ${\cal H}=\otimes_{i=1}^n{\cal H}_{l_i}$ formed
from
metaplectic representations of ${\gr sl}(2)$ with highest weights $l_i
={\b_i\over
2}+{1\over 4}$. Since these are just the functional Bethe ansatz eigenstates
for
the $[{\gr sl}(2)]^n$ Gaudin spin chain within these representations,
their completeness is equivalent to the completeness of the Bethe ansatz for
this
case.

  It may be worthwhile noting that the decomposition of the relevant Hilbert
space ${\cal H}$ into the $2^n$ irreducible subspaces formed from tensor
products of
metaplectic representations of the type  $[({1\over 4}) \oplus ({3\over 4})]^n$
suggests the presence of a supersymmetric structure involving representations
of
${\gr osp}(2,1)$ (cf. \cite{25}). Gaudin spin chains based on such
superalgebras
have been considered recently \cite{26} within the Bethe ansatz approach.
Extending the functional formulation to such superalgebras may serve to further
clarify the r\^ole of the ${\bf Z}_2^n$ invariance of the Laplacian and
the associated commuting operators on $S^{n-1}$, as well as to prove the
completeness of the Bethe ansatz eigenstates in the supersymmetric case.

  The method of solution used here may also be extended to other commuting
operators on $S^{n-1}$ arising from the $\wt{\gr sl}(2)_R$ loop algebra
approach,
including, e.g., harmonic oscillator interactions \cite{15,18,27} and
modifications of the ${\mu_i^2\over x_i^2}$ interaction terms in \HS due to
degeneracies in the space of parameters  $\{\a_1, \dots \a_n\}$ \cite{3,28}.
However, since the homogeneity operator $D$ does not necessarily commute with
the
operators in question, the joint eigenfunctions may no longer simply be
homogeneous
polynomials; in fact, they need not be polynomials at all. These questions and
further extensions of the loop algebra separation of variables method will be
addressed in future work.

\bigskip\bigskip
\noindent{\it Acknowledgements.} The authors have benefited from helpful
discussions with A. el Gradechi and E.G. Kalnins on matters related to this
work.
This research was supported in part by the Natural Sciences and
Engineering Research Council of Canada and the Fonds FCAR du Qu\'ebec.
\bigskip \bigskip
\references

1& Kuznetsov, V. B., ``Equivalence of two graphical calculi'',
{\it J. Phys. A} {\bf 25}, 6005--6026 (1992).

2& Kuznetsov, V. B., ``Quadrics on real Riemannian spaces of constant
curvature, Separation of variables and connection with Gaudin magnet'',
{\it J. Math. Phys.} {\bf 33}, 3240--3250 (1992).

3& Kalnins,  E.G.   Kuznetsov, V.B. and  Miller, W. Jr.,
 ``Quadrics on Complex Riemannian Spaces of Constant Curvature, Separation of
Variables  and the Gaudin Magnet'', {\it J. Math. Phys.} {\bf 35}, 1710--1731
(1994).

4&  Gaudin, M., ``Diagonalization d'-une classe d'hamiltoniens de spin'',
 {\it J. Physique} {\bf 37}, 1087-1098 (1976).

5&  Gaudin, M., {\it La fonction d'onde de Bethe}, Masson, Paris (1983).

6&  Adams, M.R. and Harnad,  J., ``A Generating Function Proof of the
Commutativity
of Certain 	Hamiltonian Flows'', {\it Lett. Math. Phys.} {\bf 16}, 269--272
(1988).

7&  Adams, M.R., Harnad, J. and  E. Previato,  ``Isospectral
Hamiltonian Flows in Finite and Infinite Dimensions I.
Generalised Moser Systems andMoment Maps into Loop Algebras'', {\it Commun.
Math.
Phys.} {\bf 117}, 451--500 (1988).

8&  Adams, M.R., Harnad, J. and  Hurtubise, J.,
``Dual Moment Maps to Loop Algebras'', {\it Lett. Math. Phys.} {\bf 20},
 294--308 (1990).

9&  Harnad, J., ``Isospectral Flow and Liouville-Arnold Integration in Loop
Algebras'', in: \break {\it Geometric and Quantum Methods in Integrable
Systems},
 Springer Lecture Notes in Physics  {\bf 424}, ed. G. Helminck,
Springer-Verlag,  N.Y., Heidelberg, (1993).

10&  Sklyanin, E.K., ``Separation of Variables in the Gaudin Model'',
{\it J. Sov. Math.} {\bf 47} 2473--2488 (1989).

11& Hobson,  E.W., {\it The Theory of Spherical and Ellipsoidal Harmonics},
Ch. 11,  Chelsea, New York (1965).

12& Patera, J. and  Winternitz, P., ``A new basis for the representation of
the rotation group. Lam\'e and Heun polynomials'',  {\it J. Math. Phys.}
{\bf 14}, 1130--1139 (1973).

13&  Adams, M.R., Harnad, J. and  Hurtubise, J.,
``Liouville Generating Function for Isospectral Hamiltonian
Flow in Loop Algebras'', in:  {\it Integrable and
Superintergrable Systems}, ed. B.A.  Kupershmidt, World Scientific,
Singapore (1990).

14&  Adams, M.R., Harnad, J. and   Hurtubise, J., ``Darboux
Coordinates and Liouville-Arnold Integration  in Loop Algebras'',
{\it Commun. Math. Phys.} {\bf 155}, 385-413 (1993).

15& Harnad, J. and  Winternitz, P., `` Classical and Quantum Integrable Systems
in $\wt{{\gr gl}}(2)^{+*}$ and Separation of  Variables'', preprint CRM-1921
(1993), hep-th9312035, to appear in {\it Commun. Math. Phys.} (1994).

16& Ushveridze, A.G.,  {\it Quasi-exactly Solvable Models in
Quantum Mechanics}, Ch. 5, (1994).

17&  Faddeev,  L.D. and Takhtajan, L.A. {\it Hamiltonian Methods in the Theory
of
Solitons}, \break Springer-Verlag, Berlin (1986).

18&  Macfarlane, A.J., ``The quantum Neumann model with the
potential of Rosochatius'', \break {\it Nucl. Phys.} {\bf B386}, 453-467
(1992).

19& Gagnon,  L.,  Harnad,  J., Hurtubise,  J. and  Winternitz, P.,
``Abelian Integrals and the Reduction Method for an Integrable Hamiltonian
System'', {\it J. Math. Phys.} {\bf 26}, 1605--1612 (1985).

20&  Barut, A.O. and  Girardello, L.,  ``New ``Coherent'' States Associated
with
Non-Compact Groups'', {\it Commun. Math. Phys.} {\bf 21}, 41--55 (1971).

21&  Basu, D., ``The Barut-Girardello coherent states'', {\it J. Math. Phys.}
{\bf 33 }, 114--121  (1992) .

22& Sternberg, S. and  Wolf, J.A., ``Hermitian Lie Algebras and Metaplectic
Representations'', {\it Trans. Amer. Math. Soc.} {\bf 238}, 1-43 (1978).

23&  Szeg\"o, G., {\it Orthogonal Polynomials}, Vol. {\bf XXIII},
 3rd edn.,  pp.151 ff. Amer. Math. Soc. Colloq. Publications,
Providence, R.I. (1967)

24&  Vilenkin, N.J., {\it Special Functions and Theory of Group
Representations},
Amer. Math. Soc., Providence, R.I. (1966).

25& Hughes, J.W.B.,  ``Representations of $OSP(2,1)$ and the metaplectic
representation'', \break {\it J. Math. Phys.} {\bf 22}, 245--250 (1981).

26& Brzezi\'nski, T. and   Macfarlane, A.J., ``On Integrable Models Related to
the $osp(1,2)$ Gaudin Algebra'', preprint DAMTP 93-60 (1993), hep-th9312099.

27& Babelon, O. and Talon,  M., ``Separation of variables for
the classical and quantum \break Neumann model'', {\it Nucl. Phys.} {\bf B379},
321
(1992).

28& Harnad, J. and Winternitz,  P.,  ``Integrable Systems in
$\wt{{\gr gl}}(2)^{+*}$ and Separation of  Variables II. Generalized Quadric
Coordinates'', preprint CRM (1994).

\endreferences

\end